\pdfoutput=1
\documentclass[superscriptaddress,nofootinbib,showkeys]{revtex4-2}
\usepackage[T1]{fontenc}  
\usepackage{amsmath}
\usepackage{amssymb}
\usepackage{amsfonts}
\usepackage{mathrsfs}
\usepackage{latexsym}
\usepackage{graphicx}  
\usepackage{tikzsymbols}         
\usepackage{dcolumn}  
\usepackage{color}
\usepackage{xcolor}  
\usepackage{url}          
\usepackage[english]{babel}
\usepackage[utf8]{inputenc}    
\usepackage{hyperref}  
\usepackage{mathtools}
\usepackage{natbib}
\definecolor{skyblue}{RGB}{23,111,193}
\hypersetup{
    colorlinks = true,
    linkcolor = {black},
    urlcolor = {skyblue},
    citecolor = {skyblue},
    linkbordercolor = {white},
    urlbordercolor = {white},
    citebordercolor = {white},
}

\definecolor{lime}{HTML}{A6CE39}
\DeclareRobustCommand{\orcidicon}{
	\begin{tikzpicture}
        \draw[lime, fill=lime] (0,0) circle [radius=0.2] 
        node[white] {{\fontfamily{qag}\selectfont \tiny ID}};
        \draw[white, fill=white] (-0.0625,0.095) circle [radius=0.007];
	\end{tikzpicture}
	\hspace{-2mm}
}

\foreach \x in {A, ..., Z}
{
\expandafter\xdef\csname orcid\x\endcsname{\noexpand\href{https://orcid.org/\csname orcidauthor\x\endcsname}{\noexpand\orcidicon}}
}

\begin{document}

\title{Einstein-Cartan pseudoscalaron inflation}

\author{Alessandro Di Marco\orcidA{}}
\email{alessandro.dimarco1@inaf.it}
\affiliation{Istituto Nazionale di Astrofisica,\\
Istituto di Astrofisica e Planetologia Spaziali (INAF-IAPS),\\
Via Fosso del Cavaliere, 100, 00133 Rome, Italy}

\author{Emanuele Orazi\orcidB{}}
\email{orazi.emanuele@gmail.com}
\affiliation{Escola de Ci$\hat{e}$ncia e Tecnologia and International Institute of Physics,
Federal University of Rio Grande do Norte, Campus Universitario-Lagoa Nova, Natal-RN 59078-970, Brazil}

\author{Gianfranco Pradisi\orcidC{}}
\email{gianfranco.pradisi@roma2.infn.it}
\affiliation{University of Rome ``Tor Vergata'' \\
and INFN, Sezione di Roma ``Tor Vergata'', via della Ricerca Scientifica 1, 00133 Roma, Italy}

\begin{abstract}
We study a class of early universe cosmological models based on Einstein-Cartan gravity and including a higher derivative term corresponding to a power of the Holst scalar curvature. 
The resulting effective action is basically given by General Relativity and an additional neutral pseudoscalar field (the pseudoscalaron), 
unequivocally related to the corresponding components of the torsion, that necessarily acquire a dynamics. 
The induced pseudoscalaron potential provides a realistic inflationary phase together with a very rich postinflationary epoch, resulting from the coupling of the pseudoscalaron to ordinary matter.
\end{abstract}


\maketitle
\tableofcontents

\section{Introduction}
\label{Introduction}

The theory of cosmological inflation \cite{Starobinsky:1980te, Guth:1980zm,Linde:1981mu,Albrecht:1982wi,Hawking:1981fz,Linde:1983gd}
(for reviews, see \cite{Linde:1990flp,Linde:2007fr,Olive:1989nu,Baumann:2009ds,Uzan:2015pfm})
provides a comprehensive explanation 
for the origin of the special initial conditions that support the standard Hot Big Bang (HBB) theory. 
These conditions include the flatness of the three-dimensional constant time hypersurfaces, 
the homogeneity and isotropy of the cosmic microwave background (CMB), 
the apparent absence of heavy relic particles such as magnetic monopoles, 
and the large (comoving) observable entropy. 
Additionally, inflation generates adiabatic, gaussian, and almost scale-invariant scalar metric perturbations. 
These perturbations are responsible for both matter inhomogeneities, that lead to the formation of the large-scale structures,
and the temperature anisotropies observed in CMB photons.
Furthermore, inflation also gives rise to tensor perturbations, \textit{i.e.} gravitational waves (GW), 
that could be detected if the inflationary energy scale is sufficiently high \cite{Mukhanov:1990me,Riotto:2002yw,Guzzetti:2016mkm}. 
The inflationary stage can be triggered in various ways, although the simplest one is the single-field slow-roll scenario
\cite{Steinhardt:1984jj,Liddle:1994dx}.
In this class of models, it is assumed that below the Planck scale, a single, homogeneous, neutral (spin 0), minimally coupled and canonically normalized (pseudo)scalar field $\phi$ called the \textit{inflaton}, 
plays the dominant role within the stress-energy tensor $T_{\mu\nu}$ of the Universe.
Specifically, the inflaton field is characterized by a scalar potential $V(\phi)$, known as the inflationary potential, 
which exhibits an almost plateau-like region and a global vacuum. 
Initially, the inflaton field is misaligned from the minima and slowly moves through the flat region. 
As a consequence, the potential contribution dominates over the kinetic term ($\dot{\phi}^2\ll V(\phi)$), 
and it mimicks the presence of a false vacuum or a (transient) effective cosmological constant,
whose energy density contribution, $V(\phi)\sim M^4_{inf}\ll M^4_p$,
leads to an almost de Sitter expansion of the Universe. 
Once the inflaton crosses a total distance $\Delta\phi$
\cite{Lyth:1996im,Efstathiou:2005tq,Easther:2006qu,Garcia-Bellido:2014eva,Garcia-Bellido:2014wfa,DiMarco:2017ihz}
and reaches a slow-roll-breaking value, inflation ends.
As a results, the kinetic term becomes important and
the inflaton rapidly falls down to the global vacuum, around which it begins to oscillate. 
The inflaton field should also be coupled to the degrees of freedom of the Standard Model (SM) or of its hypothetical extensions (BSM), 
enabling the transfer of energy density stored in the field condensate to the SM (or BSM) sectors. The reheating of the Universe then occurs, 
resulting in a hot plasma of SM (or BSM) particles with the required (comoving) entropy
and paving the way to the initial standard radiation dominated phase of the HBB cosmology
(see 
\cite{Albrecht:1982mp,Dolgov:1982th,Abbott:1982hn,Turner:1983he,Shtanov:1993es,DiMarco:2019czi,DiMarco:2021xzk} for perturbative reheating,
\cite{Dolgov:1989us,Traschen:1990sw,Kofman:1994rk,Shtanov:1994ce,Kofman:1997yn,Greene:1997fu,Greene:1997ge,Greene:1998nh} for nonperturbative reheating
and \cite{Bassett:2005xm,Frolov:2010sz,Allahverdi:2010xz,Amin:2014eta,Lozanov:2019jxc} for reviews).

The depicted single-field slow-roll inflationary scenario should naturally arise within a more fundamental scheme, 
where both extensions of the SM of particle physics at a scale higher than the electroweak one and of General Relativity (GR) 
are consistently coupled in a full quantum theory.  
The only available examples of such theories are (Super)strings or M-theory \cite{Polchinski:1998rq,Polchinski:1998rr}.
They cure the non-renormalizability \cite{Goroff:1985sz,Goroff:1985th} 
of GR and typically predict effective actions containing exactly additional exotic matter and gauge fields together with modified (super)gravities.
Unfortunately, a complete (realistic) top-down model is still not available starting from ten or eleven dimensions, due to the huge number of vacua and to the technical difficulties of selecting in a clean and complete way a four dimensional preferred vacuum. 
In this paper, we thus take a (bottom-up) effective action point of view (see {e.g.} \cite{Buchbinder:1992rb} for a general treatment), 
showing an example of a class of theories (developed in \cite{Pradisi:2022nmh}) 
where the inflaton comes unequivocally from geometry.  
In the framework of an Einstein-Cartan \cite{Hehl:1976kj,Shapiro:2001rz,Hammond:2002rm,Karananas:2021zkl} extension of GR 
(see also \cite{Shaposhnikov:2020frq,Shaposhnikov:2020gts,Shaposhnikov:2020aen,Langvik:2020nrs,Choudhury:2014hja,Desai:2015haa,Piani:2022gon,Piani:2023aof,He:2023vlj,Gialamas:2022xtt,Gialamas:2023emn} for recent early universe developments)
it is indeed possible to integrate out some components of the connection. 
Introducing a suitable non-linear term related to a power of the Holst curvature term \cite{Hojman:1980kv,Nelson:1980ph,Holst:1995pc}, 
one gets a dynamical pseudoscalar field (the pseudoscalaron) equipped with a potential able to drive an inflationary phase and, possibly, the subsequent reheating of the Universe. 
The pseudoscalaron would not exist in a Riemannian geometry, and can be coupled to SM (or BSM) matter.
Many other examples in the same class can be studied, giving rise to 
an interesting zoo of inflationary models, fully compatible with the current observational evidences. 
Moreover, the study of the rich post-inflationary epoch could also provide interesting windows towards an understanding of the reheating phase, leptogenesis, stochastic primordial GW background and Dark Matter. 
We will sketch a description of some of these issues, postponing a more detailed analysis to forthcoming papers.
The plan of the paper is as follows. 
In Section \ref{Einstein-Cartan gravity and the pseudoscalaron from dynamical torsion}, 
we briefly review the Palatini approach to metric-affine extensions of GR \cite{Kibble:1961ba,Hehl:1994ue,Blagojevic:2012bc,Ortin:2015hya,Baldazzi:2021kaf,Vitagliano:2010sr,Vitagliano:2013rna}, where the affine connection is promoted to be independent of the metric. 
We also review how to build a class of effective field theories where the torsion of the 
Einstein-Cartan connection is dynamical and classically equivalent to a pseudoscalar field, related to the so called Holst curvature term \cite{Hojman:1980kv,Nelson:1980ph,Holst:1995pc}. Finally, we introduce our  simple class of modified gravities, deriving the Lagrangian of the canonically normalized pseudoscalaron field. The pseudoscalaron potential is studied in detail in Section \ref{Inflationary model}, where we show how it is suited for inflation in a single-field slow-roll cosmological scenario.
In Section \ref{Inflationary phase}, 
we numerically solve the inflationary equation of motion for the pseudoscalaron evolution, computing the corresponding
Potential Slow-Roll Parameters (PSRP) and obtaining predictions for the scalar spectral index and the tensor-to-scalar ratio.
We utilize the standard COBE normalization 
to constrain both the inflationary scale and the parameters of the scalar potential.
In Section \ref{Reheating phase}, 
we provide an overview on the postinflationary phase for the whole class of models, stressing the dependence of the reheating epoch on the powers of the introduced non-linear Holst scalar curvature term. In particular, we 
discuss qualitative features of the models, leaving a quantitative deeper analysis to future publications.
Section \ref{Conclusions and Prospects}
contains our final remarks and comments, with a summary of the results and also mentioning several open perspectives.

Throughout this paper, we use natural units with $\hbar=c=1$. 
The reduced Planck mass is denoted as $M_P = 1/\sqrt{8\pi G_N}$, where $G_N$ indicates the gravitational Newton constant. 
We adopt the ``mostly plus'' spacetime signature $(-,+,+,+)$ for the four-dimensional Lorentzian spacetime metric.

\section{Einstein-Cartan gravity and the pseudoscalaron from dynamical torsion}
\label{Einstein-Cartan gravity and the pseudoscalaron from dynamical torsion}

The original Starobinsky cosmological model demonstrated that (trace anomaly) gravitational higher-order corrections
could drive a successful inflationary phase.
Subsequently, such proposal was revisited and it was soon realized that 
(i) inflation is basically controlled just by the $R^2$ correction and that 
(ii) this version is classically equivalent to GR plus an additional scalar field, the so-called \textit{scalaron}, 
canonically coupled to GR and equipped with a nontrivial potential
\cite{Teyssandier:1983zz,Whitt:1984pd,Mijic:1986iv,Mijic:1987bq,Maeda:1987xf}.
This insight was also recognized in the supergravity context \cite{Cecotti:1987sa}
and further extended,
suggesting that effective modified gravity theories with a (phenomenological) lagrangian density of the form $f(R)$
could correspond to the well known scalar-tensor theories \cite{Brans:2005ra,Sotiriou:2006hs,Fujii:2003pa}.
In an effective field theory approach that aims to describe 
the coupling of the Standard Model of particle physics to gravity, 
higher order terms also emerge quite naturally by quantum corrections\footnote{As happens in the original Starobinsky investigation.}
or in the realm of UV (quantum) completions of GR, 
like Supergravities or (Super)string/M-theory.
Moreover, such extensions of GR are better described if the connection is promoted to be an independent field with respect to the metric. 
Indeed, the choice of (metric and torsionless) Levi-Civita connection in GR is a convenient option, 
basically related to the choice of the (natural) Einstein-Hilbert action.  
A metric-affine connection differs from the Levi-Civita connection by a tensor that we dubb (as in \cite{Pradisi:2022nmh}) distorsion.  
Gravity theories based on a generic (linear) affine connection are known as ``metric-affine'' gravities \cite{Kibble:1961ba,Hehl:1994ue,Blagojevic:2012bc,Ortin:2015hya,Baldazzi:2021kaf,Vitagliano:2010sr,Vitagliano:2013rna}.
In this paper, we limit ourselves to Einstein-Cartan manifolds, where the connection has torsion but is metric compatible.  The reason is that spinor fields coupled to gravity must also be included in the matter sector.  
Following the notations and conventions in \cite{Pradisi:2022nmh}, the distorsion is defined as 
\begin{equation} 
    C_{\mu~\sigma}^{~\,\rho} \equiv {\cal A}_{\mu~\sigma}^{~\,\rho}-\Gamma_{\mu~\sigma}^{~\,\rho}, 
\end{equation}
where $ {\cal A}_{\mu~\sigma}^{~\,\rho}$ is the generic connection, while $\Gamma_{\mu~\sigma}^{~\,\rho}$ is its Levi-Civita component. Obviously, the  torsion $T_{\mu\nu\rho}$ is the antisymmetric part of the distorsion 
\begin{equation}  
    \label{torsion-distorsion}
     T_{\mu\nu\rho} \equiv C_{\mu\nu\rho}-C_{\rho\nu\mu}.
\end{equation}
The curvature associated with ${\cal A}_{\mu~\sigma}^{~\,\rho}$ is defined by 
\ \begin{equation} {\mathcal R}_{\mu\nu~~\sigma}^{~~~\rho} \equiv \partial_\mu{\cal A}_{\nu~\sigma}^{~\,\rho}-\partial_\nu{\cal A}_{\mu~\sigma}^{~\,\rho}+{\cal A}_{\mu~\lambda}^{~\,\rho}{\cal A}_{\nu~\sigma}^{~\,\lambda}-{\cal A}_{\nu~\lambda}^{~\,\rho}{\cal A}_{\mu~\sigma}^{~\,\lambda}.\end{equation}
and can be expressed in terms of $C_{\mu~\sigma}^{~\,\rho}$ as 
\begin{equation} 
    \label{FRC} 
     {\cal R}_{\mu\nu~~\sigma}^{~~~\rho}=R_{\mu\nu~~\sigma}^{~~~\rho} +D_\mu C_{\nu~\sigma}^{~\,\rho}-D_\nu C_{\mu~\sigma}^{~\,\rho}+ C_{\mu~\lambda}^{~\,\rho} C_{\nu~\sigma}^{~\,\lambda}- C_{\nu~\lambda}^{~\,\rho} C_{\mu~\sigma}^{~\,\lambda},   
\end{equation}
where $R_{\mu\nu~~\sigma}^{~~~\rho}$ is the ``standard'' (Levi-Civita) Riemann tensor. 
The curvature tensor can be contracted to provide  the usual Ricci scalar curvature 
\begin{equation} 
    {\cal R} \equiv {\cal R}_{\mu\nu}^{~~~\mu\nu}
\end{equation}
and a pseudoscalar called the Holst invariant (see \cite{Hojman:1980kv,Nelson:1980ph,Holst:1995pc})
\begin{equation} 
    {\cal R'} \equiv \varepsilon^{\mu\nu\rho\sigma}{\cal R}_{\mu\nu\rho\sigma},
\end{equation}
where $\varepsilon^{\mu\nu\rho\sigma}$ is the totally antisymmetric Levi-Civita tensor with 
$\sqrt{-g} \, \varepsilon^{0123}= 1$. Note that ${\cal R'}$ vanishes for $C_{\mu~\sigma}^{~\,\rho}=0$, namely when the connection is the distorsionless Levi-Civita one. This is the reason why in the standard formulation of GR ${\cal R'}$ is always absent.
However, it plays a prominent role in the class of theories we are going to consider, where the distorsion is dynamical. 
As shown in \cite{Pradisi:2022nmh}, they can be described using an action of the form 
\begin{equation} 
    S[g_{\mu\nu},\Phi] = \int d^4x\sqrt{-g}\left(\alpha(\Phi)  {\cal R} + \beta(\Phi)  {\cal R}' + \Delta(\Phi, {\cal R}, {\cal R}')+ \Sigma(\Phi, {\cal D}\Phi, C)    \right), 
\label{curvLag}\end{equation}
where with $\Phi$ we generically denote all fields that do not depend upon the distorsion and enter the action in combinations that respect the (global and local) symmetries present in the lagrangian, together with its scalar nature. It should be noticed that the $\alpha$ and $\beta$ functions contain (possible) non-minimal couplings to the scalar and pseudoscalar curvatures, while $\Delta$ is an arbitrary function of the indicated fields and curvatures that brings the non-linear terms.  It must be chosen appropriately (see ref. \cite{Pradisi:2022nmh} for more details) in order to guarantee that field redefinitions do not take the model back to cases where the distorsion is non-vanishing but non-dynamical.  
Finally, $\Sigma(\Phi, {\cal D}\Phi, C)$ contains the ``matter'' fields.
Notice that it depends on the distorsion through the covariant derivatives built out of the whole $\cal{A}$ connection, and thus on $C$.  

In this paper, we focus on a simple class of theories that are exactly solvable and give rise to an interesting set of inflationary models where the inflaton can be identified with a pseudoscalar field representing exactly a pseudo-scalar combination of the distorsion components, thus originating directly and unequivocally from the geometry of the underlying spacetime.  
To stress the inflationary scenario, we take $\Sigma=0$ and select $2 \alpha(\Phi)= M_{P}^2$ (thus directly the ``Einstein frame'') 
and $4 \gamma  \beta(\Phi) =M_{P}^2$ where $\gamma$ is known as the Barbero-Immirzi parameter  \cite{BarberoG:1994eia,Immirzi:1996di}. 
Moreover, we choose $\Delta$ depending solely by the pseudoscalar curvature and of the form\footnote{Some aspects of the $p=2$ case have already been discussed in \cite{Salvio:2022suk}.}
\begin{equation}
    \label{OurDelta}
    \Delta({\cal{R}'}) \, = \, \xi  \, {{\cal{R}}'}^p \ ,
\end{equation}
where $p>1$ is a real number and $\xi$ is a coupling constant with the dimensions of mass $[m]^{4-2 p}$ .
As shown in \cite{Pradisi:2022nmh}, one may introduce an auxiliary pseudoscalar field $z$, in such a way that the considered model is classically equivalent to
\begin{equation}
    S[g_{\mu\nu},z]=\int d^4x\sqrt{-g}\left[\frac{M_p^2}{2} {\cal R}+\left(\beta+\frac{\partial \Delta (z)}{\partial z }\right) {\cal R'} + \Delta(z) -z \frac{\partial \Delta (z)}{\partial z}\right] \ ,
\end{equation}
provided $\frac{\partial^2\Delta}{\partial z^2}\neq 0$.  Indeed, the equation of motion of the auxiliary fields yields $z={\cal{R}}'$, giving back the previous model on-shell.  It is now an easy algebraic exercise to decompose the distorsion into its irreducible components and to integrate it out.  Defining the quantity 
\begin{equation}
    B(z)= \frac{\beta+\frac{\partial \Delta(z)}{\partial z}}{M_P^2} \ ,
\end{equation}
it happens that its derivative sources the equations of motion of the vectorial and pseudovectorial components of the distorsion. 
In other words, on shell the action can be written as 
the sum of the Einstein-Hilbert action and the lagrangian density of the pseudoscalar field $z$, 
\begin{equation}
\label{Eframeaction}
    S[g_{\mu\nu},z] = \int d^4x\sqrt{-g}\left\{\frac{M_P^2}{2} R-K(z)\frac{(\nabla B(z))^2}{2}-V(z) \right\} , 
\end{equation}
where $R$ is the usual Levi Civita scalar curvature, the function $K(z)$ reads as
\begin{equation} 
    \label{K-B relation}
    K(z)= \frac{24 M_p^2}{[1+16 B^2(z)]} 
\end{equation}
and the potential results
\begin{equation}
    V(z) =  z \frac{\partial \Delta(z)}{\partial z} - \Delta(z) .
\end{equation}
The action in Eq.\eqref{Eframeaction} suggests that $B(z)$ brings about the (non-canonical) kinetic term related to the pseudoscalaron $z$,  that in turn certainly is not a ghost, since $K(z)$ is always positive.
As usual, a field redefinition allows to rewrite the pseudoscalar action in a canonical way.
%Moreover, 
Firstly, we have to introduce a  pseudoscalar field $\phi$ whose kinetic term is the standard one. 
This can be done letting
\begin{equation}
    \phi(z) = \int^{z} d\zeta \sqrt{K(\zeta)}.    
\end{equation}
The expression of $K(z)$ in terms of $B(z)$, allows us to establish the {\it universal} relation between the pseudoscalar field $\phi$ and $B(z)$, which holds for the whole considered class of models, {\it i.e.}

\begin{equation}
    \label{phivsz}
    \phi(z) - \phi_0 = \sqrt{\frac{3}{2}} \, M_P \, \sinh^{-1}{[4 B(z)]}, \quad 
\end{equation}
where $\phi_0$ is an integration constant.
Secondly, we need to invert the previous relation to find $z$ as a function of $\phi$, in order to find the potential $V=V[z(\phi)]$.
This procedure involves the solution of a complicated non-linear differential equation related to $\Delta(z)$ and its first derivative. 
In most cases, it is not possible to find an analytic solution. Fortunately, the simple form of our choice in Eq.\eqref{OurDelta} allows to do that, as shown in the next Section.
\section{Inflationary model}
\label{Inflationary model}

\begin{figure*}
    \begin{center}
        \includegraphics[width=15cm, height=7cm]{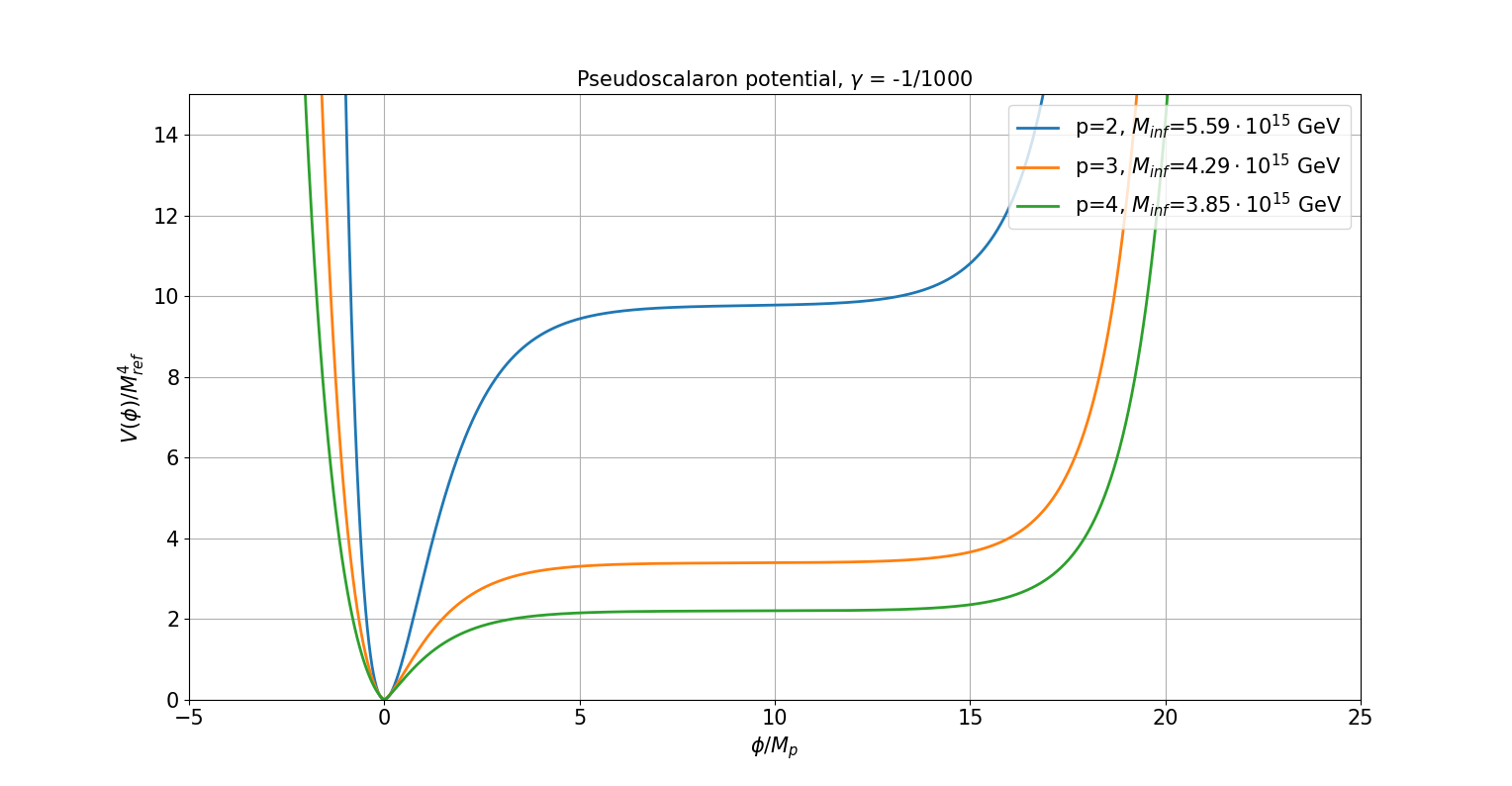}
        \includegraphics[width=15cm, height=7cm]{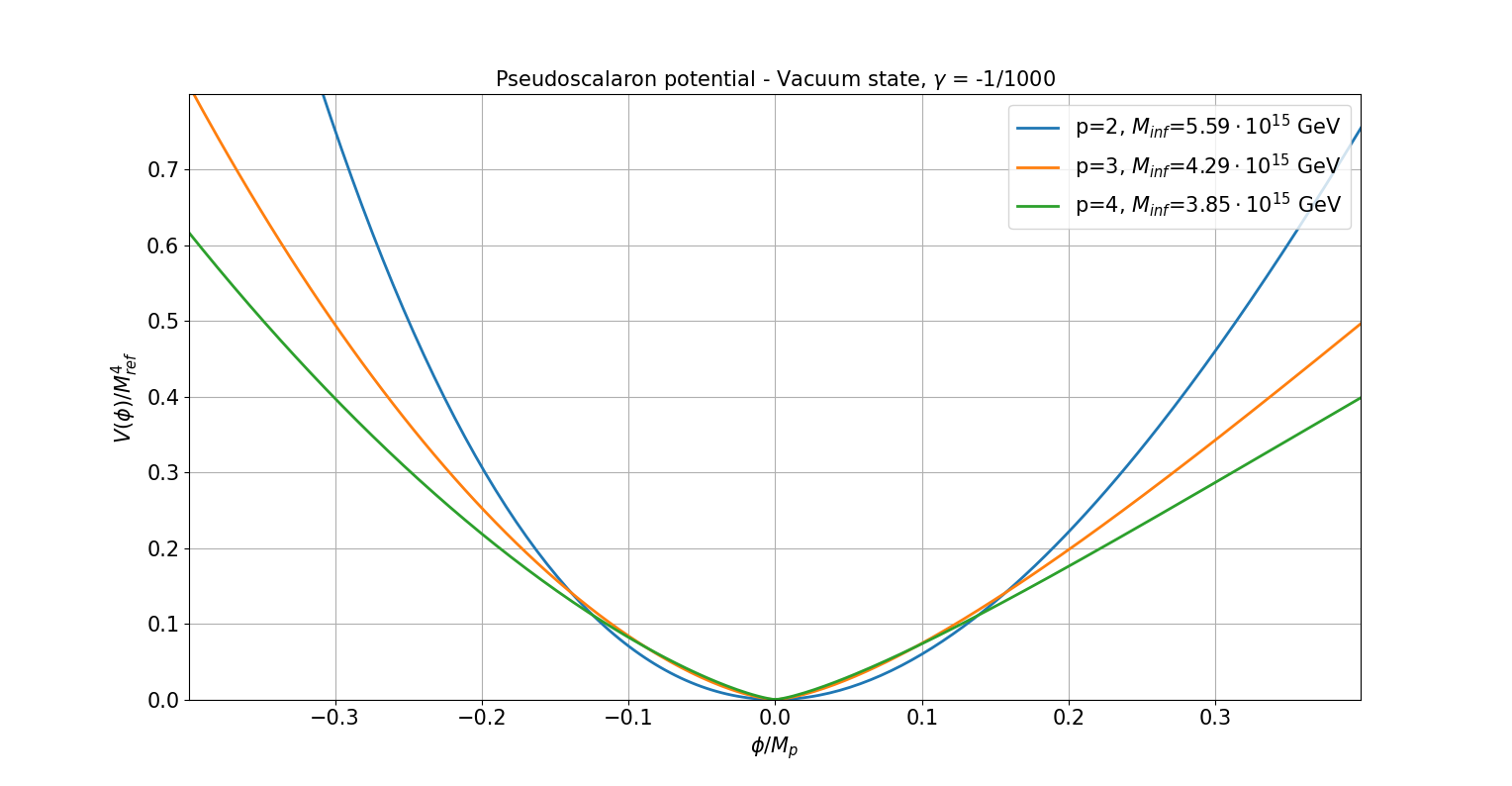}
        \caption{\it First panel: Pseudoscalaron potential normalized to a chosen reference scale $M^4_{ref}\sim 10^{62}$ GeV$^4$.
        The parameter $p$ controls the extension of the inflationary plateau. 
        In particular, smaller values of $p$ favor a potential uphill climbing for smaller values of the scalar field
        while larger values of $p$ tend to suppress the potential uphill climbing, which begins at larger field values.
        Second panel: Pseudoscalaron potential around the vacuum.
        The parameter $p$ also governs the vacuum geometry.
        It is evident that when $p$ increases, the anharmonic terms around the vacuum become significant also for $|\phi|\ll M_p$. Moreover,  for $p>2$ the potential is cuspy around the minimum.}
        \label{fig1}
    \end{center}
\end{figure*}

The assumption of Eq.\eqref{OurDelta} allows to write
the pseudoscalar sector of the action \eqref{Eframeaction} in terms of the canonically normalized field $\phi$.
Indeed, the potential can be written as
\begin{equation}
    V(z)= \xi (p-1) z^p 
\end{equation}
and Eq.\eqref{phivsz} can be explicitly inverted to give
\begin{equation}
    z^{p-1} = \frac{1}{\xi \, p} \left[ \frac{M_P^2}{4} \sinh{\left(\sqrt{\frac{2}{3}} \frac{1}{M_P} (\phi(z) - \phi_0) \right)} - \beta \right]  .
\end{equation}
Therefore the final cosmological action is
\begin{equation}
    S[g_{\mu\nu},\phi] \sim \int d^4x \sqrt{-g} \left( \frac{M^2_p}{2} R - \frac{1}{2}\partial_{\mu}\phi\partial^{\mu}\phi - V(\phi)  \right).
\label{cosmoaction}\end{equation}
Here $g$ is the determinant of the assumed Friedmann-Lemaitre-Robertson-Walker (FLRW) metric tensor $g_{\mu\nu}$, with spacetime line element given by
\begin{equation}
    ds^2 = -dt^2 + a^2(t)dl^2
\end{equation}
where $dl$ is the line element of the three-dimensional spatial subspace, $t$ is the cosmic time and $a(t)$ is the dimensionless cosmic scale factor, 
which allows to define the standard Hubble rate $H(t) = \dot{a}/a$.
Finally, the scalar potential comes out to be

\begin{equation}
    V(\phi) = \frac{p-1}{p^{p/(p-1)}}\frac{1}{\xi^{\frac{1}{p-1}}}\Bigg|\frac{M^2_p}{4}
    \sinh\left( \sqrt{\frac{2}{3}}\frac{1}{M_p} (\phi - \phi_0) \right) - \beta \Bigg|^\frac{p}{p-1} .
\end{equation}
The evolution of the inflaton as a function of the cosmic time $t$
is described by the standard Einstein-Klein-Gordon equations

\begin{align}
    \label{Einsteing-Klein-Gordon-time}
    &\Ddot{\phi}(t) + 3H(t)\dot{\phi}(t) + \frac{dV(\phi)}{d\phi} = 0, \\
    &H^2(t) = \frac{1}{3M^2_p}\left(\frac{1}{2}\dot{\phi}^2(t) + V(\phi)\right) , 
\end{align}
properly equipped with a set of initial conditions (for the field and its derivative) of the form
\begin{equation}
    \phi(t_*)=\phi_*, \quad\qquad \dot{\phi}(t_*)=\dot{\phi}_{*} .
\end{equation}
where the time $t_*$ is meant to be close enough to the horizon crossing era of observed cosmological scales, 
typically around $50/60$ $e$-folds before the end of inflation.
The vacuum expectation value (vev) $v$ of the field at the minimum of the potential is not constrained and 
typically could differ from $\phi = 0$, as happens in models with a spontaneous symmetry breaking. 
To select an inflationary model with $v=0$ 
and to interpret the pseudoscalaron $\phi$ as a particle oscillating around a flat Lorentz invariant vacuum, 
thus avoiding a Cosmological Constant term associated to the potential, 
it is customary to choose appropriately the integration constant $\phi_0$.   
In this respect, it is useful to introduce the combination
\begin{equation}
    X(\phi) = \sqrt{\frac{2}{3}} \frac{\phi}{M_p} + \theta_{\gamma}
\end{equation}
where 
\begin{equation}
    \theta_{\gamma} = \sinh^{-1}(\gamma^{-1}) ,
\end{equation}
with $\gamma$ the Barbero-Immirzi parameter.  
The condition $V(\phi)=0$ for $\phi=0$ implies that 

\begin{equation}
    \phi_0 = -\sqrt{\frac{3}{2}}M_p\sinh^{-1}{(\gamma^{-1})} .
\end{equation}
In conclusion, the scalar potential can be written in the convenient form
\begin{equation}
    V(\phi) = M^4_{inf} \, f_0(\phi) ,
\label{eq:pseudopotential}\end{equation}
where the scale of inflation can be identified with
\begin{equation}
    M^4_{inf} = \frac{p-1}{p^{p/(p-1)}}\frac{1}{\xi^{\frac{1}{p-1}}}\Bigg|\frac{M^2_p}{4\gamma}\Bigg|^\frac{p}{p-1} ,
\end{equation}
while the dependence on the field is encoded in
\begin{equation}
    f_0(\phi) = \Bigg|\gamma\sinh X(\phi)-1\Bigg|^\frac{p}{p-1} .
\end{equation}
In the spirit of an effective field theory approach, it is also useful to write the coupling strenght 
in the form $\xi = \xi_0/M_p^{2p-4}$, where $\xi_0$ is dimensionless.
The sign of the Barbero-Immirzi parameter determines the direction of the slow-roll phase.
Specifically, the slow-roll phase occurs for decreasing values of the inflaton field ({\it i.e.} $\dot{\phi}<0$)
for negative values of $\gamma$, 
while it occurs for increasing values of $\phi$
({\it i.e.} $\dot{\phi}>0$) for positive values of $\gamma$.
In Fig.\ref{fig1} both the global shape of inflationary potential and the vacuum geometry are shown, for
$\gamma = -10^{-3}$ and some chosen pairs $(p,M_{inf})$.
The Barbero-Immirzi parameter models the height and shape of the inflationary potential.
It is straightforward to show that smaller values of $\gamma$ (e.g. $\sim|10^{-2}|$) implies a higher inflation scale and a shorter plateau.
The parameter $p$ controls the extension of the inflationary plateau, the asymptotics of the potential for large field values
as well as the vacuum geometry.
Indeed, as $p$ increases, the plateau region becomes longer and the vacuum geometry becomes more and more cuspy.
It is worth noting that the shape of the inflationary potential closely resembles that of some superstring-derived inflationary scenarios  such as large volume inflation \cite{Cicoli:2008va} and in particular fibre inflation \cite{Cicoli:2008gp,Cicoli:2016xae,Cicoli:2020bao} 
(see also \cite{Baumann:2014nda} for an overview on (super)string-inspired inflationary cosmology).
For example, the expanded form of the scalar potential for $p=2$ 
and $\gamma<0$ can be written as

\begin{equation}
    V(\phi) = M^4_{inf}\left( a_0 - a_1e^{-b\phi/M_p} + a_2e^{-2b\phi/M_p} + a_3e^{b\phi/M_p} + a_4e^{2b\phi/M_p}  \right) , 
\label{ExpandedPot}\end{equation}
where $b=\sqrt{\frac{2}{3}}$ and the coefficients $a_i$ are positive and read

\begin{eqnarray}
 a_0 &= 1 - \frac{\gamma^2}{2}; \qquad  a_1 &= |\gamma| e^{-\theta_{\gamma}}; \qquad    a_2 = \frac{\gamma^2}{4}e^{-2\theta_{\gamma}} ;\\
 a_3 &= |\gamma| e^{\theta_{\gamma}}; \qquad a_4 &= \frac{\gamma^2}{4}e^{2\theta_{\gamma}} .  \nonumber
\end{eqnarray}
The mathematical difference with the fibre inflation model lays on the exponential growth of $V(\phi)$ for large field values.
In the superstring case, the divergence of the potential is just driven by a single positive exponential term, while in the present case
there are two exponential contributions related to the $a_3$ and $a_4$ coefficients.  
Exponential potentials of the form $V \sim e^{k \, \phi/M_p}$ 
are also interesting in string theory because they can provide some clues on the very onset of inflation, due to the ``climbing'' phenomenon.  
In certain orientifold models (see \cite{Angelantonj:2002ct} for a review) corresponding to $k \geq 3/2$ in $d=10$, a range whose  $d=4$ counterpart would be for $k \geq \sqrt{6}$, the dilaton can only emerge from the initial singularity ``climbing'' the exponential potential and reaching a turning point before a descent phase that can inject slow-roll inflation \cite{Dudas:2010gi, Dudas:2012vv, Kitazawa:2014dya}. 
With an exponential well, like in our case, within the interesting range  $p/(p-1)>3$ (or $p<3/2$), the behaviour near the singularity is even more complicated \cite{Condeescu:2013gaa}, with a chaotic sequence of bounces.

In the following sections, we will study the inflationary phase related to the potential in Eq. \eqref{eq:pseudopotential} for negative values of $\gamma$.

\section{Inflationary phase}
\label{Inflationary phase}

\begin{figure*}
    \begin{center}
        \includegraphics[width=15cm, height=7cm]{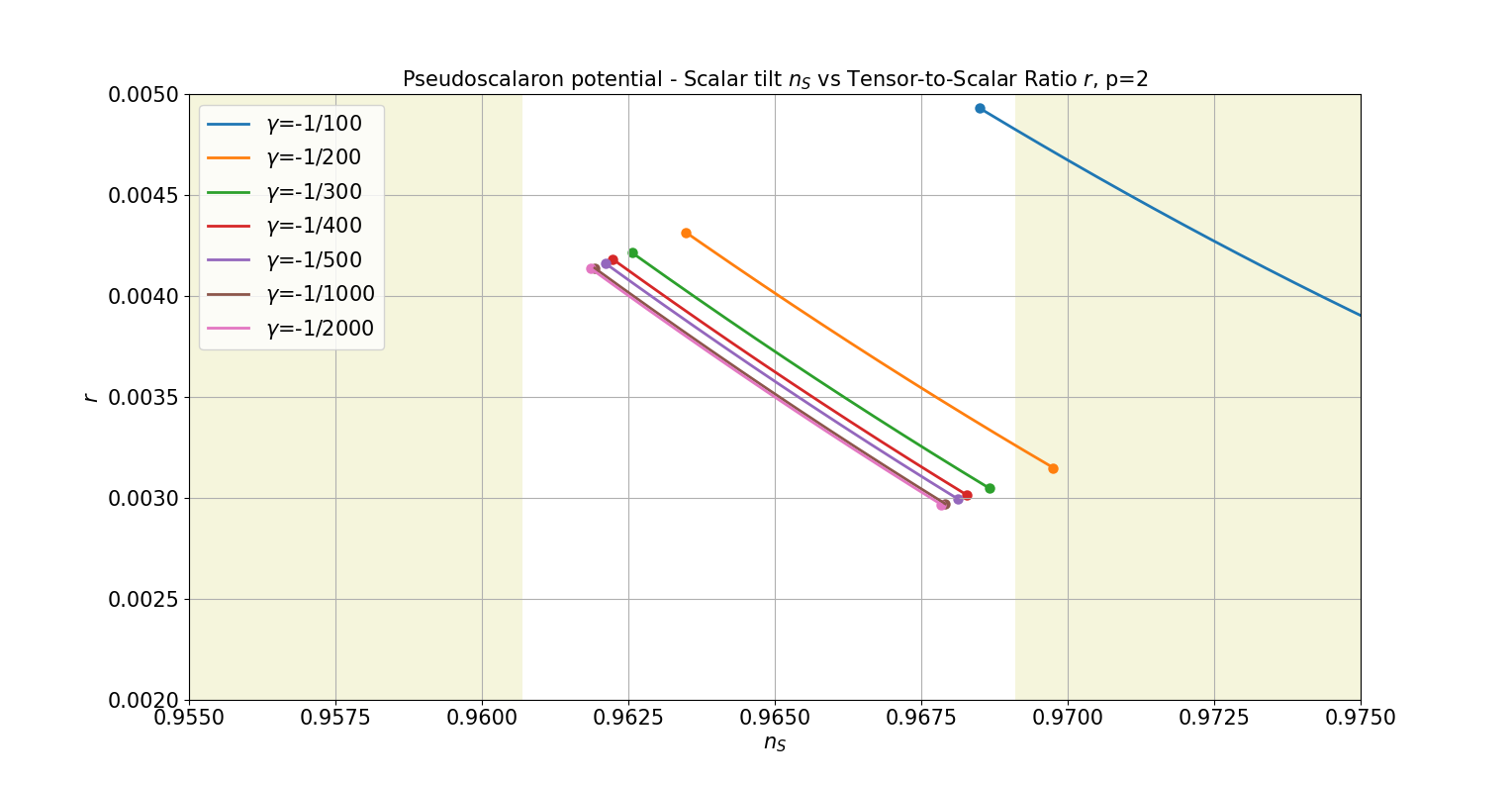}
        \includegraphics[width=15cm, height=7cm]{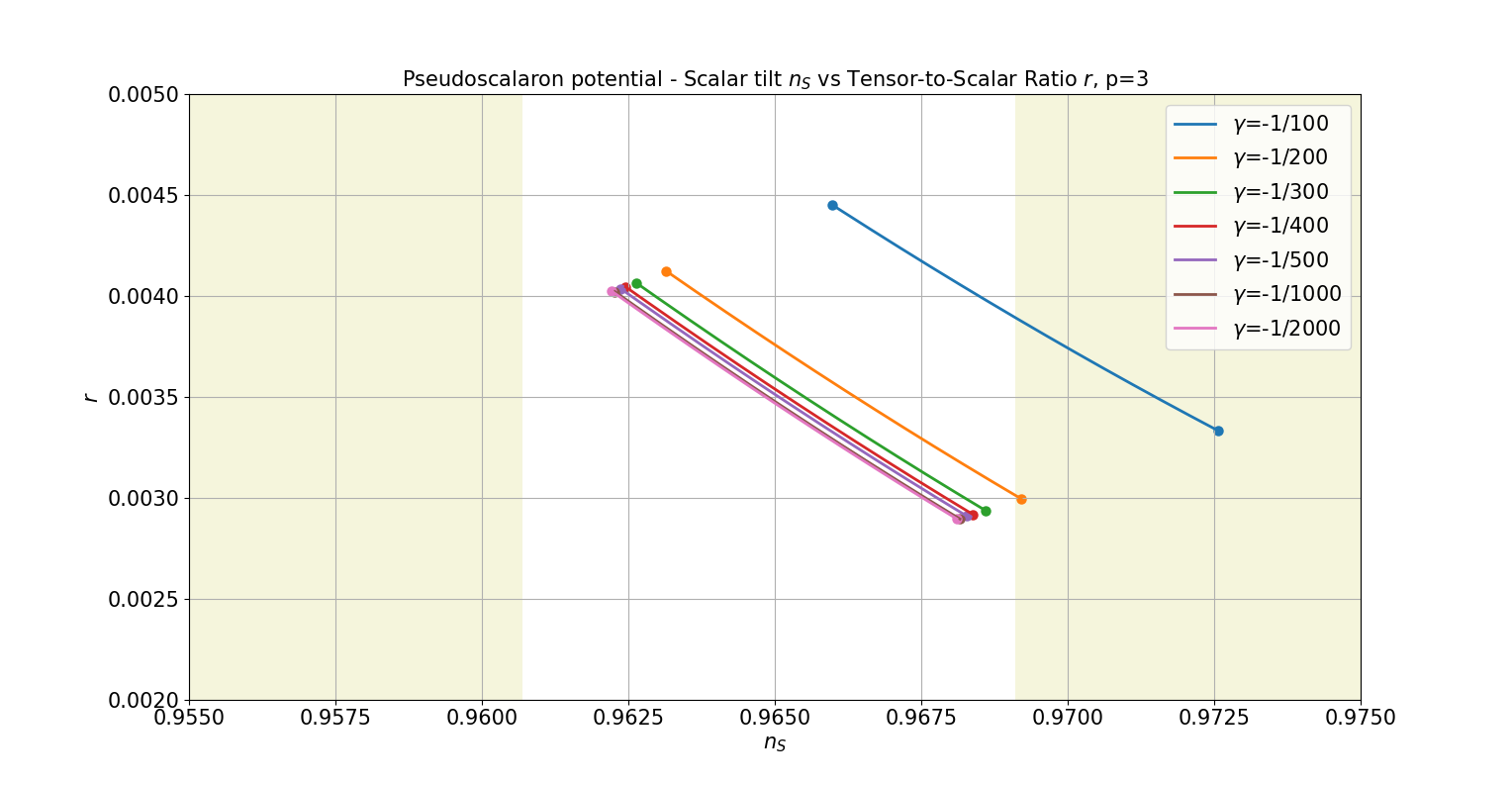}
        \includegraphics[width=15cm, height=7cm]{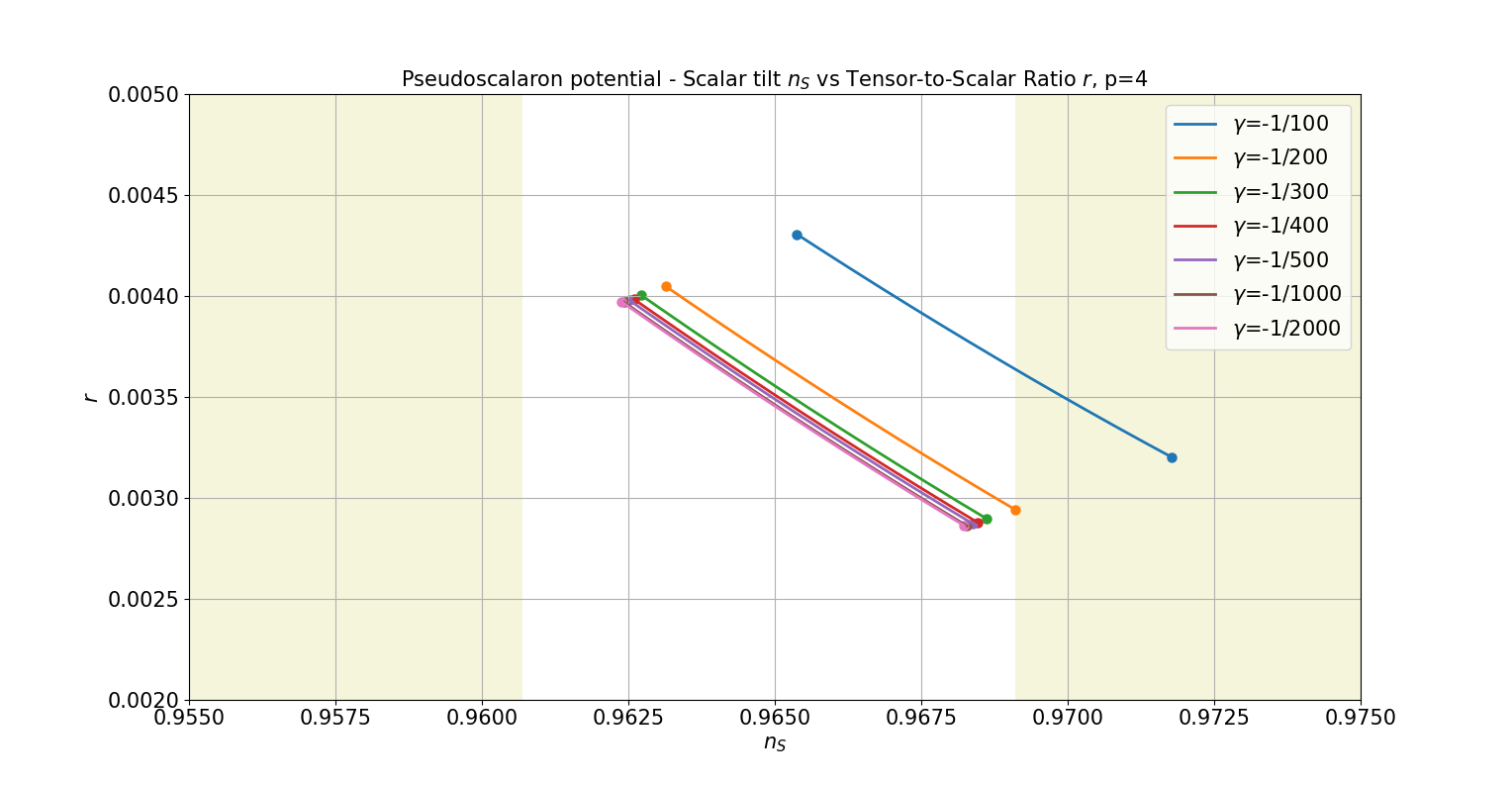}
        \caption{\it Observational constraints on pseudoscalaron inflation for three different values of the parameter $p$
        and a set of values for the Barbero-Immirzi parameter. The case $p=2$ shows that large values (in modulus) of the $\gamma$ parameter
        are heavily disfavoured while smaller values turns to be compatible with the Planck constraints. As $p$ gets larger, large values
        of the Barbero-Immirzi parameters tend to become partially compatible.}
        \label{fig2}
    \end{center}
\end{figure*}

To start the analysis of the inflationary scenario, it is
fundamental to characterize the slow-roll parameters and thus to compute the derivatives of the potential. One gets
\begin{equation}
    V'(\phi) = \frac{M^4_{inf}}{M_p}f_1(\phi) , 
\end{equation}
where
\begin{equation}
    f_1(\phi) = \dfrac{\sqrt{2} \, \gamma \,  p\, \cosh{X(\phi)}\left|\gamma\sinh{X(\phi)}-1\right|^\frac{p}{p-1}}{\sqrt{3}\left(p-1\right)\left(\gamma\sinh{X(\phi)}-1\right)} ,
\end{equation}
and
\begin{equation}
    V''(\phi) = \frac{M^4_{inf}}{M^2_p}f_2(\phi), 
\end{equation}
with
\begin{equation}
    f_2(\phi) = \frac{2 \, \gamma \, p \, \left|\gamma\sinh{X(\phi)}-1\right|^\frac{p}{p-1}}{3 (p-1)\left(\gamma\sinh{X(\phi)}-1\right)^2}
    \left(\frac{\gamma}{p-1}(p\sinh^2{X(\phi)}+1)-\sinh{X(\phi)}\right) .
\end{equation}
%With the introduced parametrization, it is possible to easily get the 
The introduced parametrization easily provides the
Potential Slow-Roll Parameters \cite{Liddle:1994dx} purely as combinations of the functions $f_i(\phi)$. 
In particular, at the leading order the first two PSRP are
\begin{equation}
    \epsilon_V(\phi) = \frac{1}{2}\left(\frac{f_1(\phi)}{f_0(\phi)}\right)^2
\end{equation}
and
\begin{equation}
    \eta_V(\phi) = \frac{f_2(\phi)}{f_0(\phi)} ,
\end{equation}
namely 
\begin{equation}
    \epsilon_V(\phi) = 
    \frac{\gamma^2}{3}\left(\frac{p}{p-1}\right)^2
    \frac{\cosh^2{X(\phi)}}{\left(\gamma\sinh{X(\phi)}-1\right)^2}
\label{epsilonV}\end{equation}
and
\begin{equation}
    \eta_V(\phi) = 
    \frac{2\, \gamma \, p}{3 (p-1)\left(\gamma\sinh{X(\phi)}-1\right)^2}
    \left(\frac{\gamma}{p-1}(p\sinh^2{X(\phi)}+1)-\sinh{X(\phi)}\right) .
\end{equation}
As usual, the PSRP allow to infer the inflationary predictions.  
The (superhorizon solutions of the) scalar and tensor power spectra in terms of the horizon crossing epoch
commonly read as
%The (scalar and tensorial) power spectra under the horizon-crossing condition are usually defined as
\begin{equation}
    P_S(k) \sim \frac{1}{8\pi^2 M^2_p}\frac{H^2}{\epsilon_V}\Bigg|_{k=aH}, \quad 
    P_T(k) \sim \frac{2}{\pi^2}\frac{H^2}{M^2_p}\Bigg|_{k=aH}
\end{equation}
%and result
and become, in terms of some horizon crossing inflaton value $\phi$
\begin{equation}
    P_S(\phi)\sim \frac{M^4_{inf}}{18 \, \pi^2 \, M^4_p \, \gamma^2} \left(\frac{p-1}{p}\right)^2
    \frac{\left(\gamma\sinh{X(\phi)}-1\right)^2}{\cosh^2{X(\phi)}}
    \Bigg|\gamma\sinh{X(\phi)}-1\Bigg|^\frac{p}{p-1}
\label{scalarpower}\end{equation}
and
\begin{equation}
    P_T(\phi) \sim \frac{2 \, M^4_{inf}}{3\, \pi^2\, M^4_p}
    \Bigg|\gamma\sinh{X(\phi)}-1\Bigg|^\frac{p}{p-1} .
\end{equation}
The related scalar spectral index and tensor-to-scalar ratio at first order in the PSRP result
\begin{align}
    n_S(\phi) &\sim 1 - 6\epsilon_V(\phi) + 2\eta_V(\phi), \\
    r(\phi) &\sim 16\epsilon_V(\phi),
\end{align}
or 
\begin{eqnarray}
    n_S(\phi) &\sim& 
    1 - \frac{2 \, \gamma^2}{3}\left(\frac{p}{p-1}\right)^2\frac{\cosh^2{X(\phi)}}{\left(\gamma\sinh{X(\phi)}-1\right)^2} \nonumber\\
    &+& \frac{4 \, \gamma \, p}{3 (p-1)\left(\gamma\sinh{X(\phi)}-1\right)^2}\left[\frac{\gamma}{p-1}\left(p\sinh^2{X(\phi}) - 1\right)-\sinh{X(\phi)}\right]
\end{eqnarray}
and
\begin{equation}
    r(\phi) \sim \frac{16}{3} \, \gamma^2\left(\frac{p}{p-1}\right)^2 \frac{\cosh^2{X(\phi)}}{\left(\gamma\sinh{X(\phi)}-1\right)^2} .
\end{equation}
The primary focus lies in computing such inflationary predictions for inflaton values corresponding to a
time interval compatible with the number of $e$-folds before the end of inflation between $60$ and $50$, as usual.
These inflaton values can be roughly or qualitatively determined by examining 
the final  portion of the slow-roll plateau region of the scalar potential $V(\phi)$.
Nevertheless, a preferred approach is to determine the inflaton values 
by exactly solving the inflationary equation of motion in Eq.\eqref{Einsteing-Klein-Gordon-time}, 
reformulated using as independent ``clock'' variable the number of $e$-folds instead of the cosmic time. One gets
\begin{equation}
    \label{eqEKGwithN}
    \frac{d^2\phi}{dN^2} + [3-\epsilon_N]\frac{d\phi}{dN} + [3-\epsilon_N]V_{eff}(\phi) = 0, 
\end{equation}
where the $\epsilon_N$ slow-roll parameter in terms of $N$ is defined as
\begin{equation}
  \epsilon_N = \frac{1}{2M^2_p}\left(\frac{d\phi}{dN}\right)^2,  
\end{equation}
and 
\begin{equation}
V_{eff}(\phi) = M_p\sqrt{2\epsilon_V(\phi)}.
\end{equation}
Eq.\eqref{eqEKGwithN} is a second-order differential equation, nonlinear with respect to the first derivative. 
It includes an effective potential term $V_{eff}$ determined by the first slow-roll parameter $\epsilon_V$. 
Unfortunately, this equation cannot be solved analytically and requires numerical integration. 
The result is an inflationary field trajectory $\phi(N)$ that provides inflationary predictions as functions of $N$, the number of $e$-folds before the end of inflation.
Before proceeding with the numerical integration, it would be interesting to explore the slow-roll limit of the equation and attempt to find a close, albeit approximate, relation between $\phi$ and $N$. In the slow-roll approximation, where the inertial term and $\epsilon_N$ are subdominant

\begin{equation}
    \frac{d^2\phi}{dN^2} \ll 1, \quad \epsilon_N \ll 1,
\end{equation}
the equation can be approximated with a linear first order ODE
\begin{equation}
    \frac{d\phi}{dN} + M_p\sqrt{2\epsilon_V(\phi)} \sim 0 ,
\end{equation}
whose integration allows to relate the number of $e$-folds before the end of inflation to $\epsilon_V(\phi)$ and to the field values by

\begin{equation}
     N(\phi,\phi_{end}) \sim \frac{1}{M_p} 
    \int_{\phi_{end}}^{\phi} d\phi' \frac{1}{\sqrt{2\epsilon_V(\phi')}}.
\end{equation}
The number $N$ depends on the value of $\phi$ at the end of inflation, defined by the condition

\begin{equation}
    \epsilon_V(\phi_{end})=1 .
\end{equation}
Using Eq.\eqref{epsilonV}, one gets the two solutions

\begin{equation}
    \frac{\phi_{end}^{\pm}}{M_p} =
    \sqrt{\frac{3}{2}}\left[\sinh^{-1}\left(
    -\frac{\gamma}{\sigma^2-\gamma^2} \pm 
    \sqrt{\frac{\gamma^2}{(\sigma^2-\gamma^2)^2} 
    - \frac{\sigma^2-1}{\sigma^2-\gamma^2}}
    \right)
    - \sinh^{-1}\left(\gamma^{-1}\right)\right] ,
\label{twophiend}\end{equation}
where 

\begin{equation}
    \sigma^2 = \frac{\gamma^2}{3}\left(\frac{p}{p-1}\right)^2 .
\end{equation}
Of course, given the shape of the pseudoscalaron potential in Fig.\ref{fig1}, 
$\phi_{end}$ must be chosen as the smaller, positive value among the two solutions in Eq.\eqref{twophiend} (since $\gamma<0$).
At the same time, the integrand of $N(\phi,\phi_{end})$ can be computed as  
\begin{equation}
    \frac{1}{\sqrt{2\epsilon_V(\phi')}} = \sqrt{\frac{3}{2}} \, \frac{p-1}{\gamma \, p}
    \left[
    \gamma\tanh{X(\phi')}  - \frac{1}{\cosh{X(\phi')}}
    \right].
\end{equation}
Hence, the slow-roll solution $N(\phi)$ results

\begin{eqnarray}
    \label{slow-roll solution}
    N(\phi,\phi_{end}) &=&
    \frac{3(p-1)}{2 p}
    \left[
    \ln{\Bigg|\cosh{X(\phi)}\Bigg|}
    - \frac{1}{\gamma}\tan^{-1}\left(\sinh{X(\phi)}\right)
    \right]\Bigg|_{\phi_{end}}^\phi\\ \nonumber
    &=&
     \frac{3(p-1)}{2 p}
    \ln{\frac{\Bigg|\cosh{X(\phi)}\Bigg|}{\Bigg| \cosh{X(\phi_{end})} \Bigg|}}
    -  \frac{3(p-1)}{2 \, \gamma \, p}
    \left[\tan^{-1}\left(\sinh{X(\phi)}\right) - 
    \tan^{-1}\left(\sinh{X(\phi_{end})}\right)\right] .
\end{eqnarray}
The first contribution just depends on the (log)ratio of two hyperbolic cosine while the second contribution
is the difference between the Gudermannian function computed at $\phi$ and $\phi_{end}$.
In general, this equation cannot be exactly inverted in terms of the scalar field.
However, it remains a very useful tool to compute pairs $(N,\phi)$ that can be used, for example
as initial conditions to trigger the dynamics described by the Eq.\eqref{eqEKGwithN}.
For example, if $\gamma\sim -10^{-3}$, then for $p=2$, $\phi(60)\sim 5.45$ while for $p=4$ one gets $\phi(60)\sim 4.99$.
Note that for values of the Barbero-Immirzi parameter $|\gamma|\gg 10^{-1}$
Eq.\eqref{slow-roll solution} can be properly approximated as

\begin{equation}
    \label{slow-roll solution approx}
    N(\phi,\phi_{end})\sim \frac{3(p-1)}{2|\gamma|p}
    \left[\tan^{-1}\left(\sinh{X(\phi)}\right) - 
    \tan^{-1}\left(\sinh{X(\phi_{end})}\right)\right] .
\end{equation}
It is possible to invert such expression and find

\begin{equation}
    X(\phi) \sim \sinh^{-1}(\tan{\omega(N)}), \quad 
    \frac{\phi(N)}{M_p} = -\sqrt{\frac{3}{2}}\theta_{\gamma} + \sqrt{\frac{3}{2}}\sinh^{-1}(\tan{\omega(N)}) ,
\end{equation}
where $\omega(N)$

\begin{equation}
    \omega(N) = \frac{2|\gamma|p}{3(p-1)}N(\phi,\phi_{end}) + \tan^{-1}\left(\sinh{X(\phi_{end})}\right) .
\end{equation}
In Fig.\ref{fig2} we provide the inflationary predictions derived through numerical integrations for $N$ in the range $\left[50,60\right]$.
In particular, we show the standard $(n_s,r)$ plane for the three cases of the parameter $p$, i.e. $p=2,3,4$
and a set of Barbero-Immirzi $\gamma$.
The white region represents the Planck constraints \cite{Planck:2018jri} for the scalar spectral index $n_s$, 
namely $n_s=0.9649 \pm 0.0042$ at $68\%$ of confidence level (CL).
The BICEP/Keck upper limit on the tensor-to-scalar ratio \cite{BICEP:2021xfz}, \textit{i.e}
$r_{0.05}<0.036$ at $95\%$ CL, is not visible.
The computation shows that as the parameter $p$ decreases, the inflationary predictions shift towards the central region
and become more and more compatible with the Planck constraints.
This means that both the scalar spectral index and the tensor-to-scalar ratio tend to decrease for a fixed value of $\gamma$.
At the same time, the $n_s$-$r$ curves tend to cluster together as $\gamma$ becomes larger (in modulus).
For example, if $N\sim 55$ and $\gamma\sim -10^{-3}$ then $n_s\sim 0.9650$ and $r\sim 0.0035$.
On the other hand, if one considers the case $p=2$ the scenario $\gamma\sim -10^{-2}$ is practically disfavoured by the standard observations
while for $\gamma\sim 0.5\times 10^{-2}$ the compatibility with the observations is almost recovered.
Instead, the general predicted amplitude of GW via the $r$ parameter is of the order of $\sim 10^{-3}$ and it is aligned
with the plethora of models typically inspired by supersymmetry, supergravity or superstring theories.

Let us also comment about the scale of inflation, namely the value of $M_{inf}$. 
By requiring that the scalar power spectrum of Eq.\eqref{scalarpower} coincides with the COBE measured $P_S^{COBE}\sim 2\times 10^{-9}$, we obtain
\begin{equation}
    \frac{M_{inf}}{M_p}\sim \left[ 
                             8 \, \pi^2 \, \gamma^2 \, P_S^{COBE}
                             \left(\frac{p}{p-1}\right)^2
                             \frac{\cosh^2{X(\phi)}}{\left(\gamma\sinh{X(\phi)}-1\right)^2} 
                             \frac{1}{\Bigg|\gamma\sinh{X(\phi)}-1\Bigg|^{p/(p-1)}}
                             \right]^{1/4}.
\end{equation}
The previous expression, fixed $\gamma$, $p$ and $N$ (or in other words the associated value of $\phi$), 
allows to constrain the coupling parameter $\xi$ that can be inferred to be

\begin{equation}
    \xi\sim \left[ 
            \frac{p-1}{p^\frac{p}{p-1}} 
            \frac{1}{M^4_{inf}}
            \Bigg|\frac{M^2_p}{4\gamma}\Bigg|^{p/(p-1)} 
            \right]^{(p-1)} .
\end{equation}
It is important to note that such a coupling constant
depends in a non-trivial way on the ratio of the Planck mass on a certain power of the Barbero-Immirzi parameter $\gamma$.
In Tab. I, we present various scenarios for the inflationary energy scale $M_{inf}$ and the magnitude of the
dimensionless coupling constant $\xi_0 = \xi M_p^{2p-4}$.
Notably, it can be observed that the coupling strength is significantly suppressed for $p>2$, 
despite a slight variation in the inflation scale. 
This outcome directly stems from the power $(p-1)$: 
as $p>2$, the dominant contribution becomes $M^{-4}_{inf}$, leading to a damping effect on the value of $\xi$.

\begin{table*}
\label{tab1}
    \begin{ruledtabular}
    \centering
        \begin{tabular}{ccccccc}
                                        &     \multicolumn{3}{c}{$\gamma=-10^{-2}$}                 &                        &\multicolumn{1}{c}{$\gamma=-10^{-3}$}                     \\ 
         $p$      &     $\phi(60)/M_p$  &     $M_{inf}$ [GeV]           &  $\xi_0$ [GeV]$^{2p-4}$   &      $\phi(60)/M_p$    &      $M_{inf}$ [GeV]         &      $\xi_0$ [GeV]$^{2p-4}$\\ \hline
         $2$      &     $5.53$          &     $1.31\times 10^{16}$      &  $5.38\times 10^{9}$      &       $5.45$           &      $5.59\times 10^{15}$    &      $1.60\times 10^{13}$\\
         $3$      &     $5.16$          &     $1.10\times 10^{16}$      &  $6.74\times 10^{19}$     &       $5.12$           &      $4.29\times 10^{15}$    &      $1.16\times 10^{26}$\\
         $4$      &     $5.02$          &     $1.01\times 10^{16}$      &  $1.19\times 10^{30}$     &       $4.99$           &      $3.85\times 10^{15}$    &      $1.28\times 10^{39}$\\
        \end{tabular} 
    \end{ruledtabular}
\caption{Constraints on inflationary scale and coupling parameter $\xi_0 = \xi M_p^{2p-4}$ for some selected values 
of the Barbero-Immirzi parameter and the power $p$. 
The simplest case $p=2$ provides a very large coupling $\xi$ while for smaller values, the coupling values is higly suppressed.
The inflationary scale could be easily of the order of $\sim 10^{16}$ GeV.}
\end{table*}

\section{Postinflationary phase}
\label{Reheating phase}
The inflationary phase typically dilutes all the preexisting energy and entropy densities,
providing a very cold and almost empty Universe.
However, as the pseudoscalaron reaches the corresponding slow-roll ending condition value $\phi_{end}$, 
it rapidly relaxes in the global vacuum and starts to oscillate, paving the way for the reheat of the Universe.
The properties of the scalar field oscillations depend on the local vacuum geometry and are obviously damped by the background Universe expansion.
In pseudoscalaron inflation there are basically two possibile scenarios.  
Let us start by the first one, already partially discussed in \cite{Salvio:2022suk} and relative to the case in which the potential in Eq.\eqref{eq:pseudopotential} exhibits a smooth global minimum. 
It happens in the case with $p=2$, where the pseudoscalaron potential can be expanded around the (Lorentz invariant) vacuum as
\begin{equation}
    V(\phi)\sim \frac{1}{2}m_{\phi}^2\phi^2 + \frac{g_{\phi}}{3}\phi^3 + \frac{\lambda_{\phi}}{4}\phi^4 + ...
\end{equation}
The pseudoscalaron acquires both a finite mass $m^2_{\phi}=d^2V(\phi)/d\phi^2|_{\phi=0}$ and higher order self-coupling contributions.
The mass term naturally overwhelms the higher-order interactions and 
results larger than the Hubble rate $H$ ($m_{\phi}\gg H$).
As a consequence, the (homogeneous) pseudoscalaron field begins to oscillate with a frequency $\omega_{\phi}$ compatible with the mass $m_{\phi}$ 
and with a period of oscillation $t_{\phi}$ shorter than the Hubble time scale $t_{\phi}\sim m_{\phi}^{-1}\ll H^{-1}$. 
In other words, a Hubble expansion tipically contains a huge number of field oscillations. 
At this stage of the Universe evolution, the system can be thought of as a condensate of a large number of heavy pseudoscalaron quanta
with mass $m_{\phi}$ and zero momentum. Thus, the inflaton can  be described in terms of a perfect fluid 
with an averaged Equation-of-State \cite{Turner:1983he,Shtanov:1993es,DiMarco:2019czi} 
$w_{\phi}=(n-2)/(n+2)\sim 0$, since $V(\phi)\sim \phi^n$ with $n=2$.
In order to properly reheat the Universe and to generate the corresponding (comoving) entropy density, 
the pseudoscalar inflaton must couple to the particles of the SM/BSM sectors, producing them in the open decay channels.
Several decay products can appear in the $\Sigma(\Phi, {\cal D}\Phi, C)$ part of the Lagrangian of Eq.\eqref{curvLag}. For instance, one may have spinor fields 
minimally coupled to Einstein-Cartan gravity. Additional effective couplings among matter fields, that respect the symmetries of the total lagrangian and that can be generated by quantum corrections are also allowed. Among them, 
quite interesting are effective non-minimal couplings of scalar fields to the scalar curvatures as well as effective couplings of (abelian) gauge fields to the vectorial and pseudovectorial components of the distorsion.
The former can give rise, for instance, to decays of the inflaton directly to the Higgs SM sector \cite{Salvio:2022suk}. 
The latter allow mixings of abelian gauge fields to vector or pseudovector fields, 
paving the way to an interpretation of these new dynamical components of the distorsion as dark photons \cite{Pradisi:2022nmh}. 
The interactions with ordinary (or dark) matter are very model-dependent, 
being connected to the way they mix to the ``visible'' photon and to the form of the gravity sector. 
Some effective interaction terms of the expansion can be guessed, at least at the lowest order. 
For instance, a non-minimally coupled additional scalar field $\chi$ can be considered. 
It means to have in Eq.\eqref{curvLag} the $\alpha$ term no-longer constant but function of $\chi$. 
At the lowest order it results 
\begin{equation}
   \alpha(\chi) = \frac{M_P^2}{2} + g_{\chi} \chi^2 , 
\label{nonmincoup}\end{equation}
where $g_{\chi}$ is a  dimensionless coupling constant. It results in an interaction term in the effective Lagrangian of the form 
\begin{equation}
    \mathcal{L}_{int}^{\chi} \sim \frac{c_{\phi\chi\chi}}{M_P} \, \chi \, \partial_{\mu}\phi \, \partial^{\mu}\chi .
\end{equation}
Analogously, for a Dirac fermion $\psi$ minimally coupled the interaction term is of the form
 \begin{equation}
     \mathcal{L}_{int}^{\psi} \sim \frac{c_{\phi\psi\psi}}{M_P} \, \partial_{\mu}\phi\, \bar\psi \, \gamma^5 \, \gamma^{\mu} \, \psi .
 \end{equation}
The exact coefficients can be obtained by integrating out the non-dynamical part of the connection, as before. 
Notice that the coupling strenghts $c_{\phi i i}$ are typically functions of the Barbero-Immirzi parameter $\gamma$.
The inflaton energy (density) conversion process is characterized by a perturbative phase, sometimes preceded by a preliminary non-perturbative regime, although the details strongly depend on the model parameters as well as on the couplings $c_{\phi i i}$.
In the case of a negligible or absent nonperturbative regime, the inflaton decay and the related particle production 
is just a single-particle (non collective) process, basically driven by the decay rates of the inflaton to the daughter particles $\Gamma_{\phi\rightarrow i_1, ..., i_n}$.
Such decay amplitudes typically depend on the inflaton mass, while the total decay amplitude is the sum over all the possible channels. %their sum provides
As is well known, the impact of this process on the inflaton oscillations about the vacuum can be modeled by introducing 
a phenomenological friction term $\sim \Gamma_{\phi}\dot{\phi}$ in the inflaton equation of motion, 
that represents a further source of damping in addition to the standard Hubble friction.
It is important to stress that the computed total decay amplitude $\Gamma_{\phi}$ is crucial to establish
the out-of-equilibrium decay of the pressureless pseudoscalaron (average) energy density $\rho_{\phi}$ and the corresponding growing of
the (SM/BSM) radiation one $\rho_{r}$, via the usual Einstein-Boltzmann equations 
\cite{Turner:1983he,Shtanov:1993es,DiMarco:2019czi,DiMarco:2021xzk}
\begin{eqnarray}
    \dot{\rho}_{\phi} + 3H\gamma_{\phi}\rho_{\phi} &=& -\gamma_{\phi}\Gamma_{\phi}\rho_{\phi}\\
    \dot{\rho}_{r} + 4H\rho_{r} &=& + \gamma_{\phi}\Gamma_{\phi}\rho_{\phi}
\end{eqnarray}
where $\gamma_{\phi} = 1 + w_{\phi}$ and $w_{\phi}=0$ in the present case. The Hubble parameter is defined by the Freedman equation
\begin{equation}
    H^2 = \frac{1}{3M^2_p}\left(\rho_{\phi} + \rho_r \right) ,
\end{equation}
while the initial conditions are
\begin{equation}
    \rho_{\phi}(t_{end})=\rho(\phi_{end}), \quad \rho_{r}(t_{end}) = 0 .
\end{equation}
This evolution is characterized by a final reheating temperature that notoriously scales as $T_{reh}\sim \sqrt{M_p\Gamma_{\phi}}$.
As qualitatively discussed in \cite{Salvio:2022suk}, 
the decay amplitude of the pseudoscalaron to fermion is proportional to $\sim m_{\phi}m^2_{\psi}/M^2_p$. Therefore a quite large fermion mass (\textit{e.g} $m_{\psi}\lesssim m_{\phi}$)
is required in order to reach a reheating temperature well above the electoweak scale. 
The SM does not contain such heavy fermions and an interesting possibility 
would be to insert in the matter sector, in addition to the SM fields,  sterile Right-Handed Neutrinos (RHN), that naturally possess such large masses.   
Moreover, RHN could also work as the crucial ingredients to get nonthermal leptogenesis in this scenario.  
Clearly, a decay channel of the pseudoscalaron to non-minimally coupled scalars like that in Eq.\eqref{nonmincoup} 
would be sufficient to reheat the Universe in a more efficient way to the needed temperature. 
Indeed, the decay rate is independent of $m_{\chi}$ and proportional to $\sim m^3_{\phi}/M^2_p$.  
As noted in \cite{Salvio:2022suk}, the scalar field $\chi$ could naturally be identified with the SM Higgs. 
Thus, the presence of only the distorsion in the gravity sector is sufficient to get inflation and reheating, 
without the necessity of additional BSM degrees of freedom.  

Adding a minimally coupled scalar is also a possibility. 
However, there is no direct interaction with the pseudoscalaron, since the distorsion does not enter the covariant kinetic term of the additional scalar. Rather, the two scalars would naturally mix giving rise, likely, to a multifield inflation.

All the details missed or just mentioned in this brief description will be presented in a separate publication, 
where a quantitative and complete description of perturbative reheating and Leptogenesis in the proposed scenario will be given.

Let us now briefly discuss the second class of scenarios, related to the case where $p>2$.  
As one can immediately recognize from Eq.\eqref{eq:pseudopotential}, the global minimum around $\phi=0$ is not smooth, being of the form 

\begin{equation}
  V(\phi)\sim  M^4_{inf} \, \left| \sqrt{\frac{2(1 + \gamma^2)}{3}}\,  \frac{\phi}{M_p} \right|^{\frac{p}{p-1}}  .
\end{equation}
Therefore, the scalar potential exhibits a cuspy behaviour.  
Similar potentials occur in many models, like in $k$-inflation \cite{Armendariz-Picon:1999hyi} and in string theory, 
\textit{e.g.} in axion monodromy models and in flux compactifications 
\cite{Silverstein:2008sg,McAllister:2008hb,Marchesano:2014mla,Landete:2017amp}.
What happens is the following: first of all, the field oscillates around the minimum, but clearly the oscillations cease to be harmonic
and the (would be) mass term is divergent.
Fig.3 shows an example of pseudoscalaron oscillation in the usual Minkowski spacetime (upper panel) and 
in the realistic expanding postinflationary Universe (lower panel).
In the first case, the pseudoscalaron starts from a reference plateau field value and falls in the global vacuum.
Here, it exhibits an asymmetric oscillation that is quite regular in the standard (almost) quadratic scenario, $p=2$.
However, as $p>2$,  the cuspy potential leads to a nontrivial oscillation characterized by a natural damping.
In the second case, the Hubble friction plays a crucial role in suppressing the oscillation structures of the previous scenario, 
although one can always appreciate the difference between the quadratic and cuspy cases.
\begin{figure*}
    \begin{center}
        \includegraphics[width=15cm, height=7cm]{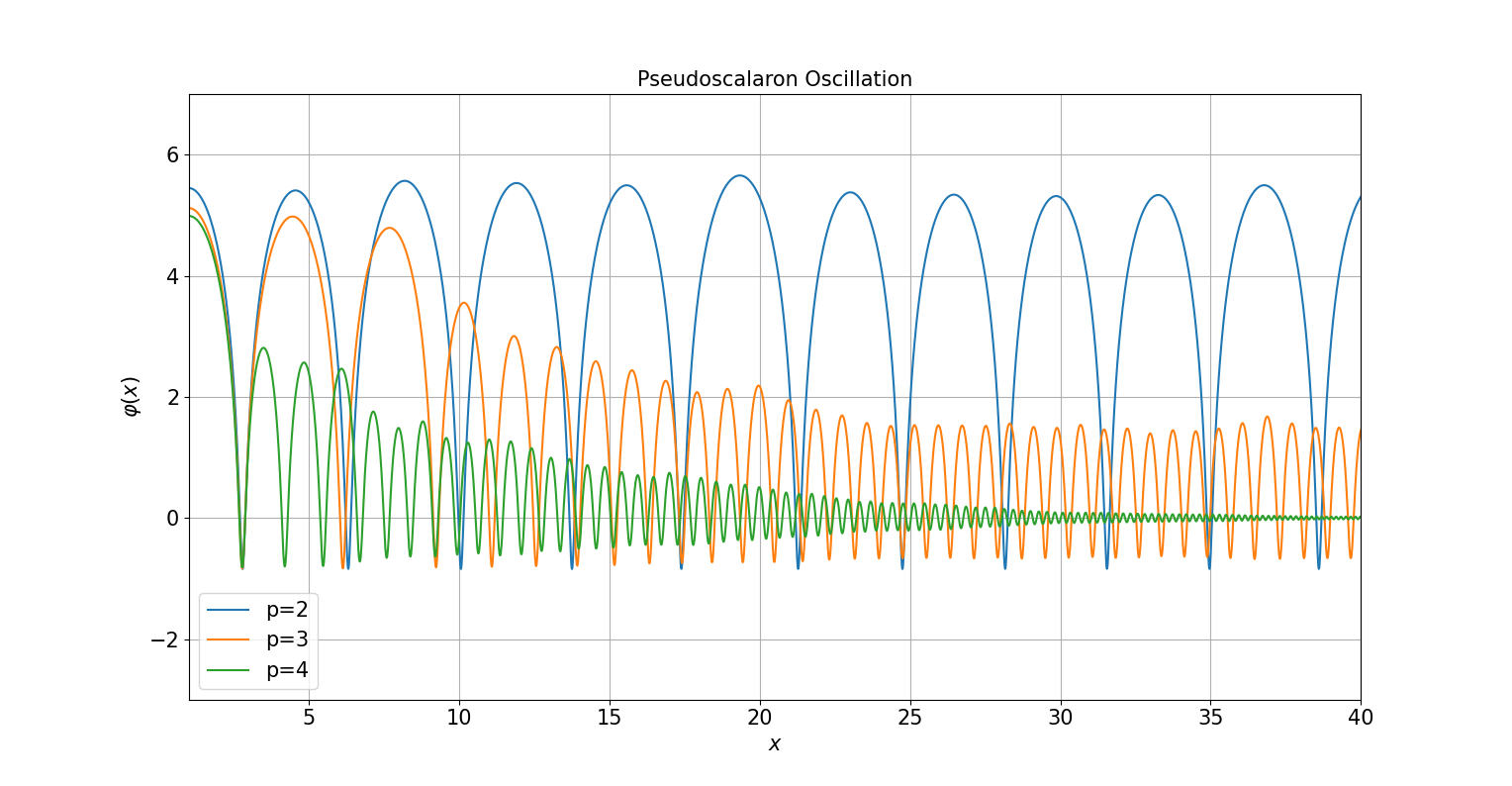}
        \includegraphics[width=15cm, height=7cm]{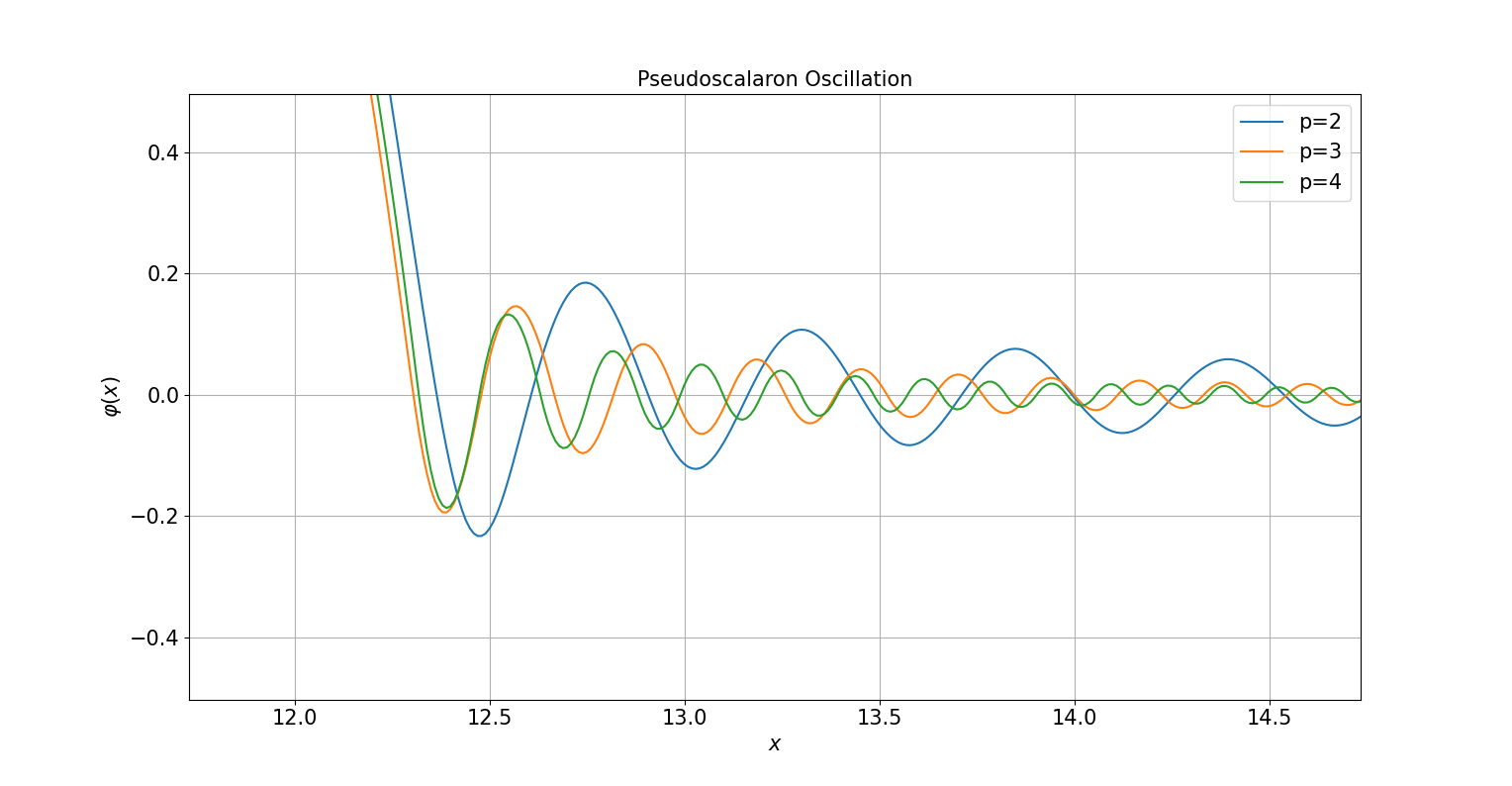}
        \caption{\it Prototype pseudoscalaron oscillation in a Minkowski spacetime (upper panel) 
        and in the more realistic postinflationary Universe (lower panel). 
        The decay of the pseudoscalaron to daughter particles is neglected.}
        \label{fig3}
    \end{center}
\end{figure*}
In the $p>2$ framework, the pseudoscalaron can still be treated as a perfect fluid that, however, is equipped 
with a highly nontrivial EoS given by 

\begin{equation}
    w_{\phi} = \frac{n-2}{n+2}\Big|_{n = p/(p-1)} = \frac{2-p}{3p-2} ,
\end{equation}
This quantity is generically different from zero when $p\ne 2$ and in the limit of $p\gg 2$, the EoS parameter takes the value $-1/3$, 
mimicking the same effect of a linear potential $V(\phi)\sim |\phi|$ around the vacuum. 
Certainly, the inflaton is still able to transfer energy to the decay products. 
Indeed, as shown in \cite{Liu:2017hua,Liu:2018rrt}, cuspy potentials naturally lead to a so-called preheating phase, 
where the (highly efficient) inflaton energy density conversion process is 
realized through an exponential growth of (the occupation number of) the pseudoscalaron decay product modes.  
In other words, collective phenomena lead to a very fast reheating of the Universe because of Bose condensation effects triggered 
by a decay product phase space that is sufficiently populated.  
Preheating in the form of parametric resonance can also be treated when the involved couplings are large (see, for instance,  \cite{Kofman:1994rk}).
In addition, preheating in cuspy potentials can also give rise to a copious formation of oscillons 
that source an additional stochastic background of primordial gravitational waves (GW) \cite{Liu:2017hua,Liu:2018rrt}.
Such GW peaks could be visible 
in the future observing run of the LIGO-Virgo experiment \cite{LIGOScientific:2017zlf}.
To summarize, the post-reheating phase of the considered models with $p>2$ is quite rich and interesting and it deserves, in our opinion, a deeper and more complete analysis. 
Again, we will report specific and detailed predictions on preheating (and related GW generation) in separate, forthcoming publications.

\section{Conclusions and Prospects}\label{Conclusions and Prospects}

Fundamental theories containing (extensions of) the SM coupled to gravity, like superstrings or M-theory, typically
cure the non-renormalizability \cite{Goroff:1985sz,Goroff:1985th} of GR 
and predict (Einstein-Cartan) modified gravities coupled to SM or BSM fields, 
together with a plethora of additional sectors. 
Pure Einstein-Hilbert GR is typically accompanied by quantum corrections, due to string loops and higher derivative corrections, 
related to massive string states.
In a bottom-up approach, it is possible to include corrections in an effective action consisting of a power series in curvature terms 
weighted by corresponding powers of the Planck mass.  
In this paper, we have discussed a class of models (a subclass of those introduced in \cite{Pradisi:2022nmh}) 
where the corrections to GR are unequivocally linked to the geometry of the (non-Riemannian) spacetime. 
In particular, we have considered Einstein-Cartan gravities with dynamical distorsion and higher derivative terms given by powers of the so-called Holst scalar curvature that, as known, vanishes in a Riemannian geometry.  
The resulting effective field theory is equivalent to GR coupled to a pseudoscalar field 
that can naturally drive a single-field slow-roll inflationary phase
fully compatible with the current observational evidences.
In addition, compatible couplings to ordinary matter provide a highly non-trivial reheating phase
that depends on the power $p$ of the (non-linear) Holst term (see Eq.\eqref{OurDelta}).
Indeed for $p=2$ the potential is smooth around the minimum, giving rise to a standard field oscillation
that can source both a preheating phase and/or a (subsequent) perturbative reheating phase
driven by the decay rates of the inflaton to SM fields or hypothetical BSM fields.
On the other hand, for $p>2$ the potential is cuspy around the minimum and the standard, almost quadratic oscillation phenomenon, is absent.
The inflaton energy density conversion can be realized via a rich preheating phase with a possible production of GW via oscillon dynamics. 
Moreover, the coupling of the pseudoscalaron to heavy RHN
and their resulting decay can pave the way for nonthermal leptogenesis and the generation of the observed baryon asymmetry of the Universe.  
We have described in detail the class of involved models and their inflationary slow-roll phase.  
We have also depicted, in a qualitative way, the postinflationary phase, stressing the involved and peculiar interesting properties.
The richness of the postinflationary phase requires and deserves a careful and deeper analysis.  
We leave it to separate, forthcoming, publications \cite{nuovo1, nuovo2}. 
Finally, It would be challenging to discover string theory setups where similar potentials are realized and give rise to the ``climbing'' phenomenon, 
as explained at the end of Section \ref{Inflationary model}. 
It would also be important to search for a new class of Einstein-Cartan gravities that generalize those proposed 
in this paper to models that are scale-invariant at any coupling, along the lines of \cite{Shaposhnikov:2020frq}
as well as \cite{Karananas:2021gco,Olmo:2022ops}.

\acknowledgments
We would like to thank A. Sagnotti and A. Salvio for many interesting and illuminating discussions.
A.D.M. has been supported by the G4S$\_$2.0 project, developed under the auspices of the Italian Space Agency (ASI) within the frame of the Bando
Premiale CI-COT-2018–085 with co-participation of the
Italian Institute for Astrophysics (INAF) and the Politecnico di Torino (POLITO). The research of E.O. is partially supported by the ``National Council for Scientific and Technological Development - CNPq'' - Brazil.

\nocite{*}

\bibliographystyle{apsrev4-2}
%-------Newcommands for hyper-bibliography (journal and arxiv)
\newcommand{\journal}[2]{\href{http://dx.doi.org/#1}{#2}}
\newcommand{\arxiv}[2]{\href{http://arxiv.org/abs/#1}{[arxiv:#1 #2]}}

\end{document}